\documentclass[11pt,letterpaper]{article}

\usepackage{amssymb}
\newcounter{bla}

\usepackage{graphicx}
\usepackage{dcolumn}
\usepackage{bm}
\usepackage{booktabs} 

\begin{document}

\title{\Large Relativistic and Electromagnetic Molecular Dynamics Simulations for a Carbon--Gold Nanotube Accelerator}

\author{Motohiko Tanaka, and Masakatsu Murakami{\small $^{1}$}\\
{\sl {\small Graduate School of Engineering, Chubu University, Kasugai  487-8501, Japan}}\\
\sl {\small $^{1}$Institute of Laser Engineering, Osaka University, Suita 565-0871, Japan}}

\maketitle

\begin{abstract}
Relativistic molecular dynamics are described in ultra-high temperature and MeV energy behaviors. In strongly-coupled systems, the Coulomb electrostatic field is collected in the infinite space, and the electromagnetic fields are added in the coordinate space. Separation of the electromagnetic and electrostatic electric fields is a good approximation for short time periods. For a numerical application, a nanotube accelerator under an $E \times B$ pulse is studied. Positive ions are accelerated in the parallel direction, whereas the electrons proceed in the perpendicular direction. Rapid expansion to infinite space and short-range electromagnetic radiation cooperate for large intensities. At $10^{22}\rm{W/cm}^{2}$, pulsation oscillations for gold and carbon ions flare up and electrons acquire the relativistic velocities. 
They are observed in relativistic molecular dynamics simulation.  
\end{abstract}

\noindent
{\small 
Keywords: molecular dynamics simulation, Coulomb potential, 
separation of longitudinal and transverse electric fields, 
parallel computing, nanotube accelerator}

\noindent
{\it Computer Physics Communications. 
DOI: 10.1016/j.cpc.2019.03.012}

\section{\label{sec1}Introduction}

Ion accelerations driven by ultraintense and ultrashort laser pulses have been studied over the past two decades. Positive ion beams have wide applications such as tomography, cancer therapy~\cite{thra1}, compact neutron sources~\cite{neutr1}, and ion-driven fast ignition for medicine~\cite{medic1}. To practically use accelerated ions, it is crucial to produce high-quality beams, which are monoenergetic and collimated~\cite{phys-pl,phys-rl,nature}. Laser-driven ion acceleration has been reviewed in the literature~\cite{review-lia}.

In laser-ion acceleration schemes, the generated ion energies are expected to be of MeV-order with 10's of fs.
A molecular dynamics simulation suggested that a nanometer-sized nanotube can generate a quasi-monochromatic and directional beam of protons and carbon ions in the MeV energy range~\cite{ref01}. Additionally, a gold coating can be chemically adsorbed to the monolayer of carbon atoms, which comprise the carbon nanotube~\cite{goldion}. Low-Z materials such as hydrogen and carbon are fully ionized, while materials such as gold have mid-range ionization states $Z\le25$ at laser intensities of $\le 10^{18}\rm{W/cm}^2$~\cite{ref10}. Increasing the irradiated laser intensity by several orders of magnitude will be an interesting challenge as both the energy and energy efficiency must be increased above $10^{18}\rm{W/cm}^2$.

On the numerical simulations, the carbon nanotube accelerator should utilize molecular dynamics to maintain the initial crystal state, which is then rapidly superseded in the MeV main phase.
The strongly-coupled system is defined by $\Gamma = e^{2}/\epsilon ak_{B}T > 1$~\cite{Frenkel,McCam}, where $e$ is a charge, $\epsilon$ is a dielectric constant, $a$ is an inter-particle spacing, and $k_{B}T$ is thermal energy. 
The simulation in such systems is calculated with electrostatic forces, where the periodic Coulomb field is by the use of the Ewald method for a large system at high accuracy~\cite{Deserno}. 
The forces are $exp(-(\gamma r)^2)/r$ in the short range, while the charge density with aliasing sums is convolved by a Fourier transform for the long range, with $\gamma$ an inverse length parameter. 
The microwave heating of water shows that the ice state is crystallized and not melted below 273 K~\cite{mwave}. 
The other polymer fields are involved in the charge inversion and electrophoresis phenomena~\cite{chginv,chginv2,chginv3}, DNA translocation in the nanopore~\cite{DNApore}, and macromolecules of polyampholyte~\cite{polyamp1}. 

In plasma physics, the electromagnetic simulations are treated in coordinate space on top of the finite-element method~\cite{AIP,FEM}. 
The electromagnetic fields are calculated for high-beta plasmas, but the electrostatic field must be solved in coordinate space to avoid accumulation of aliasing errors~\cite{brack,Macro0,LFE}. They are applied to space physics~\cite{Macro1,cmr,Asym}. 
The codes are recently developed in robust moving mesh methods of time-dependent partial differential equations~\cite{MMesh,MMesh2}.
Computation in Livermore National Laboratory is centered in many areas which are inertial confinement of laser fusion and computational mathematics on partial differential equations~\cite{Liverm}.
For the present carbon nanotube accelerator, however, the $\Gamma$ parameter is highly variable at $\Gamma > 1$ for the initial phase and $\Gamma \ll 1$ for the expanded plasma in an unbound space. 
All particles in the infinite volume must be calculated by Coulomb potential field which was previously done~\cite{ref01a}.

Solid-state plasma wakefield accelerator was proposed in two decades after 1980's~\cite{accel-1, accel-2}.
Laser pulse with a very tight focus has a focal spot size of a few $\mu \rm{m}$. The system size under consideration is larger than the Debye length~\cite{accel-11,accel-12}. For these parameters, particle-in-cell method is commonly used~\cite{accel-4,accel-13,accel-14}.
On the other hand, it is within our technical feasibility of nanotubes to assume such resolution as Angstrom to $\rm{nm}$. Our focus is on the dynamics of laser-matter (plasma) interaction of a single or few nanotubes.
Since the nanotube accelerator is difficult to solve in short ranges, one has to resort to molecular dynamics (particle-to-particle) simulation~\cite{ref01}.

In the aforementioned study, a carbon nanotube was used in the relativistic molecular dynamics simulation. The simulation was used in the electrostatic approximation at that time. 
Fortunately, molecular dynamics simulations of today's technology can utilize real mass and charge particles with electromagnetic effects. 
The scalings of these ions and electrons can be addressed without additional adjustments. 

The rest of this paper is organized as follows. In Sec.\ref{sec2}, the methodology of molecular dynamics for relativistic and electromagnetic field simulations is described. A concise procedure, which separates the Coulomb potential and transverse electric fields, is a good approximation of small-scale phenomena. 
The Courant condition for electromagnetic fields is discussed in Sec.\ref{sec2.2}.
Section \ref{sec3} describes the numerical application of the molecular dynamics simulation code. Protons, carbons, golds, and electrons are used at their real mass ratios, where the applied $E\times B$ Gaussian pulse in the time and space is given as $(\omega t -k_{y}y)$. The carbon nanotube is accelerated in the parallel and perpendicular directions for ions and electrons, respectively. Multi-nanotubes will be presented in Sec.3.2. 
The energy efficiency is about 1\% at $10^{22}\rm{W/cm}^2$. Finally, Sec.\ref{sec4} summarizes the paper. 

The international system of units will be shown in Appendix A.
Parallel computing by the MPI technique for molecular dynamics simulation will be discussed in Appendix B.

\section{\label{sec2}Methodology of Relativistic Molecular Dynamics Simulations}
\subsection{\label{sec2.1}Motion and Maxwell Field Equations}

Solving the Newton equation of motion for position $\vec{r}(t)$ and velocity $\vec{v}(t)$ can realize time development in a system. Since particle positions of $N$ particles are located in infinite space, the charged particles interactions must be highly accurate.
It is contrary to using only the numerical grids in the coordinate space. 
The Coulomb potential in the infinite space is adopted in the equation of motion, which is written as
\begin{eqnarray}
\Phi_{Coul}(\vec{r}_{i})= \sum_{j=1}^{N} q_{i} q_{j} /r_{ij}. 
\label{Coulomb}
\end{eqnarray}
Here, $q_{i}$ is the $i^{th}$ particle charge, $r_{i}$ is the position in the parallel ($z$) and lateral directions ($x$ and $y$). The vector $\vec{r}_{ij} = \vec{r}_{i} - \vec{r}_{j}$ is the displacement between the $i^{th}$ and $j^{th}$ particles. The sum of all possible combinations of particles must be specified.

In nanometer crystals in general, particles have strong short-range interactions. These interactions can be represented by the Lennard-Jones potential, which is expressed as 
\begin{eqnarray}
\Phi (r_{i},r_{j}) &=& \epsilon_{ij} [(\sigma_{ij}/r_{ij})^{12} - 2(\sigma_{ij}/r_{ij})^{6}].
\label{binary-coll}
\end{eqnarray} 
The short-range potential is repulsive if the distance between the two particles is small, but is attractive if the particle distance is larger than the equilibrium distance. The potential for carbon pairs has an equilibrium at $\sigma_{CC}= 1.421$\AA\ and $\epsilon_{CC}= 4 \rm{eV}$ ($1 \rm{eV}= 1.602 \times 10^{-12} \rm{erg}$). Hence, a solid six-membered network of nanotubes is created. The other ions are weakly dependent, and their potentials are assumed to be null, 
$\epsilon_{AuAu}= 0$. Actually, the depth of the carbon potential is about $T\approx 4 \times 10^{4} \rm{K}$. Consequently, the short-range potential works to form a nanotube crystal at the initial state. 
The electrons and ions can be heated up by the laser pulse quickly during the rising edge of the pulse.
However, one starts from such an initial crystal nanotube, where all the atoms are ionized at a temperature of 10 eV.
Within the 10-\AA\ pairs, the Coulomb and Lennard-Jones interactions are calculated at a small time step $\Delta t$, while the interactions beyond that range are kept at 5 $\Delta t$.

The electromagnetic interactions are calculated by a set of Maxwell equations in rectangular coordinates.  
Usually, the retarded potential is used for the special relativity~\cite{Landau, Heras}. Although this potential is difficult to solve for a system with a large number of particles, such as $N > 10^5$ particles, the large Coulomb potential field and the transverse electric field are separated. 
%
The dynamic range of the proton acceleration emitted from a 100-nm-long nanotube during $\cong$ 10 fs is of the same order of the target itself, and it is substantially shorter than the applied (Ti-sapphire) laser wavelength of 800 nm. 
Also in our problem, where the Debye length is significantly longer than the system size, substantial amount of electrons are blown off in the early stage. The massive ions remaining there are then accelerated by their own Coulomb repulsive force up to several to a few tens of MeV within tens of fs. Under such non-relativistic acceleration 
regime of the protons, the radiation loss are negligibly small.

Finite coordinate space grids are used to define the 
electric field $\vec{E}(\vec{r},t)$ and the magnetic field $\vec{B}(\vec{r},t)$. 
The electromagnetic fields are determined by the Maxwell equation. Round differential derivatives of space $\nabla \times$ and time $\partial/\partial t$ are the centered differential scheme, which is given by
\begin{eqnarray}
(1/c)\partial \vec{B}/\partial t &=& - \nabla \times \vec{E} 
\label{ddd.eq} \\
(1/c)\partial \vec{E}/\partial t &=& \nabla \times \vec{B} 
- (4 \pi/c) \sum_{i=1}^{N} q_{i} \vec{v}_{i} S(\vec{r}-\vec{r_{i}}), 
\label{ddd.ampere} \\
\nabla \cdot \vec{B}= 0,  
\label{ddd.divB} 
\end{eqnarray}
with $\vec{B}= \vec{H}$ in the CGS system of plasmas.
%
It is important that the divergence electric field $\nabla \cdot \vec{E} $ must be corrected because of aliasing errors of numerical grids~\cite{brack, Macro0, LFE}.
Namely, the uncorrected electric field that is deviated from the  Gauss law, $\nabla \cdot \vec{\breve{E}} \ne 4 \pi \rho$, is adjusted by,
\begin{eqnarray}
- \nabla^2 \delta \phi (\vec{r})= 4 \pi \rho (\vec{r}) -
\nabla \cdot \vec{\breve{E}}(\vec{r}, t).
\label{ETcorre1}
\end{eqnarray}  
The corrected part $\nabla \delta \phi$, in support of the Fourier space, is written as,
\begin{eqnarray}
\vec{E}(\vec{r}, t)= \vec{\breve{E}}(\vec{r}, t) 
- \nabla \delta \phi (\vec{r}),
\label{ETcorre2}
\end{eqnarray}
which is used in Eqs.(\ref{ddd.eq}) and (\ref{ddd.ampere}).

The transverse electric field $\vec{E}_{T}(\vec{r},t)$ and the  longitudinal electric field $\vec{E}_{L}(\vec{r})$ are separated in order to use the Coulomb forces, instead of $\vec{E}_{L}$, for the resolution accuracy in Eqs.(\ref{eq_motion}) and (\ref{eq_motion2}) below. The longitudinal electric field is first solved in the coordinate space, 
\begin{eqnarray}
\nabla \cdot \vec{E}_{L}(\vec{r})= 4 \pi \rho(\vec{r}) = 4 \pi \sum_{i=1}^{N} q_{i}S(\vec{r} -\vec{r_{i}}). 
\label{EL1} 
\end{eqnarray}
The summation is taken over particles and $S(\vec{r})$ is prorated to nearby grids (the shape function is $S(\vec{0})=1$ and $S(\vec{r})\rightarrow 0$ when $|\vec{r}| \gg \Delta z$). 
The real-odd function of the fast Fourier transform is used~\cite{FFTW} as $\vec{E}_{L}(\vec{r})= -\nabla \Phi_{L}(\vec{r})$ and $\Phi_{L}(\vec{r})  \propto \sin(k_{x}x)\sin(k_{y}y)\sin(k_{z}z)$. 
The transverse electric field is then separated from the longitudinal electric field,
\begin{eqnarray}
\vec{E}_{T}(\vec{r}, t)= \vec{E}(\vec{r},t) - \vec{E}_{L}(\vec{r}).
\label{EL2}
\end{eqnarray}  
%
The transverse electric field $\vec{E}_{T}$ and the Coulomb force field are used in the equation of motion. 

The time step for the $\vec{E}_{T}$ 
field is $t= t^{n+1/2}$, where $t^{\alpha}$ stands for the time level $t=t^{\alpha}$.
After this separation, the Coulomb forces and electromagnetic fields are used in Eq.(\ref{eq_motion}) below.
The electromagnetic fields are solved in the coordinate, which are due to the available meshes from computer resources.
The space operator of three dimensions for the electric field, for example, is given by
\begin{eqnarray}
\nabla \times \vec{E} =  
(\partial E_{y}/\partial z - \partial E_{z}/\partial y, 
 \partial E_{z}/\partial x - \partial E_{x}/\partial z, \nonumber \\
 \hspace{6cm}
 \partial E_{x}/\partial y - \partial E_{y}/\partial x) .
\label{curl}
\end{eqnarray}
The time advance is $t^{n-1/2} \rightarrow t^{n+1/2}$ for the magnetic field $\vec{B}$, and the transverse electric field $\vec{E}_{T}$ is advanced by $t^{n} \rightarrow t^{n+1}$.

The relativistic equation of motion to define momentum $\vec{p}_{i}$, position $\vec{r}_{i}$, and velocity $\vec{v}_{i}$, is written by the centered ordinary differential scheme as, 
\begin{eqnarray}
d \vec{p}_{i}/dt &=& - \nabla \sum_{j=1}^{N} [q_{i}q_{j} /r_{ij} + \Phi (r_{i},r_{j})] 
 \nonumber \\[-0.3cm]
 && \hspace{1.7cm} 
 + q_{i} [\vec{E}_{T}(\vec{r}_{i},t) +(1/c)\vec{v}_{i} \times \vec{B}(\vec{r}_{i},t)],
\label{eq_motion} \\
d \vec{r}_{i}/dt &=& \vec{v}_{i}, \ \ 
\vec{p}_{i}= m_{i}\vec{v}_{i}/\sqrt{1 - (\vec{v}_{i}/c)^2}.
\label{eq_motion2}
\end{eqnarray}
%
The electrostatic terms in the first bracket, namely Eq.(\ref{Coulomb}) and Eq.(\ref{binary-coll}), and the electromagnetic terms in the second bracket of Eq.(\ref{eq_motion}) are used here. 
Here, $m_{i}$ is the mass of the $i^{th}$ particle, $c$ is the speed of light, the symbol $d/dt$ is the full time derivative, and $\nabla$ is an operator $(\partial/\partial x, \partial/\partial y, \partial/\partial z)$. 
The time advance for the momentum $\vec{p}$ is made at $t^{n} \rightarrow t^{n+1}$, and that for the position $\vec{r}$ is by a half-time staggered at $t^{n+1/2} \rightarrow t^{n+3/2}$.
This cycle shown in Eqs.(\ref{ddd.eq})-(\ref{eq_motion2}) is repeated in a next time step.

The coordinate space has the meshes centered at the origin. The grids of $5 \ \AA$ have spacing $(500)^{2} \times 1000$ \AA$^{3}$ in $\vec{E}_{T}(\vec{r}_{i},t)$ and $\vec{B}(\vec{r}_{i},t)$ in the $x, y$ and $z$ directions for $< 10^{21}\rm{W/cm}^{2}$, but must have $2.5 \ \AA$  spacing at $(500)^{2} \times 1500$ \AA$^{3}$ for large intensity $10^{22}\rm{W/cm}^{2}$. 
In the external regime, other components are null except for $(-\nabla \Phi )$ in Eq.(\ref{Coulomb}) and the drift terms $(E_{z,0}, B_{x,0})(\vec{r},t)$. The Gaussian pulse with an 800-nm wavelength ($2.67\times 10^{-15} s$) is induced from the negative $y$ direction toward the origin at the speed of light. It decays in $\Delta Y=2.8 \rm{\mu m}$  (3.5 wave periods) in the $y$ direction, namely $3.5 \times 800 \rm{nm}$, and $\Delta Z=0.1 \rm{\mu m}$ in the $x,z$ directions,
\begin{eqnarray}
 E_{z,0}(\vec{r}, t) &=& E_{0} sin(\omega t -k_{y}y) *f_{n}, \nonumber \\
 B_{x,0}(\vec{r}, t) &=& B_{0} cos(\omega t -k_{y}y) *f_{n}, 
\label{EBpulse}
\end{eqnarray}
and the slope $f_{n}$ is
\begin{eqnarray}
 f_{n} = exp[-( (y-p_{xyz})^2/\Delta Y^2 +(x^2+z^2)/\Delta Z^2)].
\end{eqnarray} 
Here, the oscillation frequency is $\omega = ck_{y}$, and $E_{0}=B_{0}$. The slopes are $p_{xyz}= (-3 +t/2.67 \times 10^{-15}) \times 800 \rm{nm}$. The time step is $\Delta t=5 \times 10^{-19} s$. 
The laser intensity $5 \times 10^{17}\rm{W/cm}^2$ corresponds to the electric field $1.46 \times 10^{12} \rm{V/m}$ in the MKSA unit, which is $4.58 \times 10^{7} \rm{statV/cm}$ in the CGS unit. 
%

\subsection{\label{sec2.2}Ionization state and Courant condition}

\begin{figure}
\centering
  \includegraphics[width=6.5cm,clip]{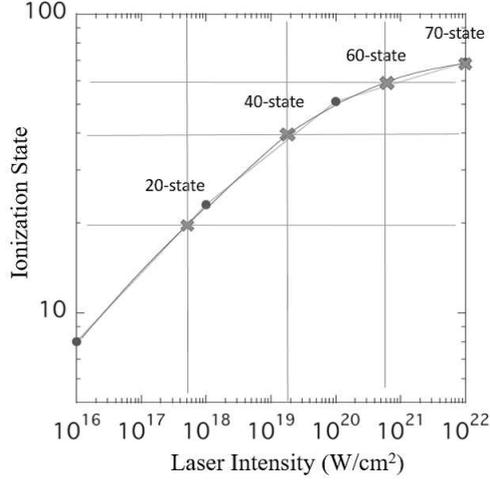}
\caption{
Laser intensity and ionization state of gold ions for $10^{16} - 10^{22} 
\rm{W/cm}^2$.}
\label{Laser-intesity}
\end{figure}

The values studied in this simulation range between $10^{16}$ to $10^{22}\rm{W/cm}^2$. The carbon nanotube on the surface of the two-dimensional cylinder surface is about 15 nm (radial, diameter) and 30 nm (height, diameter). The gold ions are set by 2-\AA\ positions outside of the accompanying carbon pairs. Moreover, inside the inner cylinder, the bulk, which has a 12-nm diameter and 27-nm full length (height), is uniformly filled with protons and electrons. 
The proton and carbon ions are fully ionized due to the high laser intensity. On the other hand, the gold ions are partially ionized, depending on the laser intensity~\cite{ref10}, as depicted in Fig.\ref{Laser-intesity}. Each run is assigned an ionization state, $Z_{is}$= 20, 40, 60, and 70, which are designated as Run A1, A2, A3, and A4, respectively.

The plasma contains: (i) carbon ions with six ionization states and 55,000 particles, (ii) gold ions as described above with 55,000 particles, (iii) protons with a single ionization state and 10,000 particles, and (iv) electrons with a negative single ionization state. The charge neutrality condition determines the number of electrons. Particles ($4.1 \times 10^{5}$) are assigned as protons, carbons, golds, and electrons. The initial temperature of the particles is equal to null. 
Run A1 to A3 are executed as long runs with $t=$ 40-60 fs with $L_{z}=\pm 500 \ \AA$.  
Run A4 arises a rapid expansion which subsides and lasts $t \le$ 20 fs, and full electromagnetic fields require the space accuracy.
It is executed by $\Delta z=2.5 \ \AA$ with $L_{z}=\pm 750 \AA$.
%
A multi-nanotube of a double or triple cages (without protons) are executed as $5 \ \AA$ accuracy for Run B, and three arrays of a multi-nanotube with double layers inside are presented in Run C of Sec.\ref{sec3.2}.

To minimize the number of electrons, the electrons are disguised such that electrons are lumped up as super-electrons with a super-charge and mass of $e_{\alpha}=-\alpha e$ and $m_{\alpha}=\alpha m_{e}$. One super electron with $e_{\alpha}=-6e$ is assigned to a carbon ion. For gold, one super electron with $e_{\alpha}=-5e$ is assigned to one-forth of the gold ions for Run A1. In the same token, one super electron with $e_{\alpha}=-10e$ is assigned to one-fourth of the gold ions for Run A2, one electron $e_{\alpha}=-15e$ is assigned to one-fourth of the gold ions for Run A3, and one super electron with $e_{\alpha}=-(70/4)e$ is assigned to one-fourth of the gold ions for Run A4. A positive ion, namely a proton, carbon, or gold, is used as the independent normal particle.

The Courant-Friedrichs-Lewy condition must be satisfied for the numerical simulation code to be stable for electromagnetic waves~\cite{CFL}. The time step $\Delta t$ and grid interval $\Delta z$ must be greater than the speed of light, 
\begin{equation}
\Delta z/c \Delta t  > 1,
\end{equation}
otherwise the simulation blows up immediately for the explicit code. 
The most demanding condition for Run A4 with 2.5 \AA \ should be  
\begin{eqnarray}
\Delta z/c \Delta t &=& 2.5 \times 10^{-8}\rm{cm}/ c \times 5 \times 10^{-19}\rm{sec} = 1.67,
\end{eqnarray}
which satisfies the aforementioned condition.

\begin{table}
\caption{The kinetic energy of carbons, golds, protons and electrons (MeV, average) for the laser intensity in Runs A1 to A4 ($\rm{W/cm}^2$). 
The energy efficiency of golds over electrons (\%) is also listed.}
\label{Table-1}
  \centering 
  \vspace{0.1cm}
  \begin{tabular}{ccccccc} \toprule
  Series & laser intensity & C  & Au   & Proton  & electron & Efficiency \hspace{0.3cm} \\ \hline 
  A1 & $5.0 \times 10^{17}$ & 0.32 & 0.59 & 0.085 & 0.036 & 69\% \hspace{0.3cm}
\\
  A2 & $1.7 \times 10^{19}$ & 0.76 & 6.0   &  0.21 &   0.53 & 25\% \hspace{0.3cm} 
\\
  A3 & $6.0 \times 10^{20}$ &  0.82 & 13.9 &  0.25 &  6.1 & 3.1\% \hspace{0.3cm}
\\
  A4 &	 $1.0 \times 10^{22}$ & 1.9 & 21.9 &  0.73 & 37.3 & 0.78 \%
\hspace{0.3cm} \\ \toprule
  \end{tabular} 
\end{table}

\begin{figure} 
\centering
\includegraphics[width=9.0cm,clip]{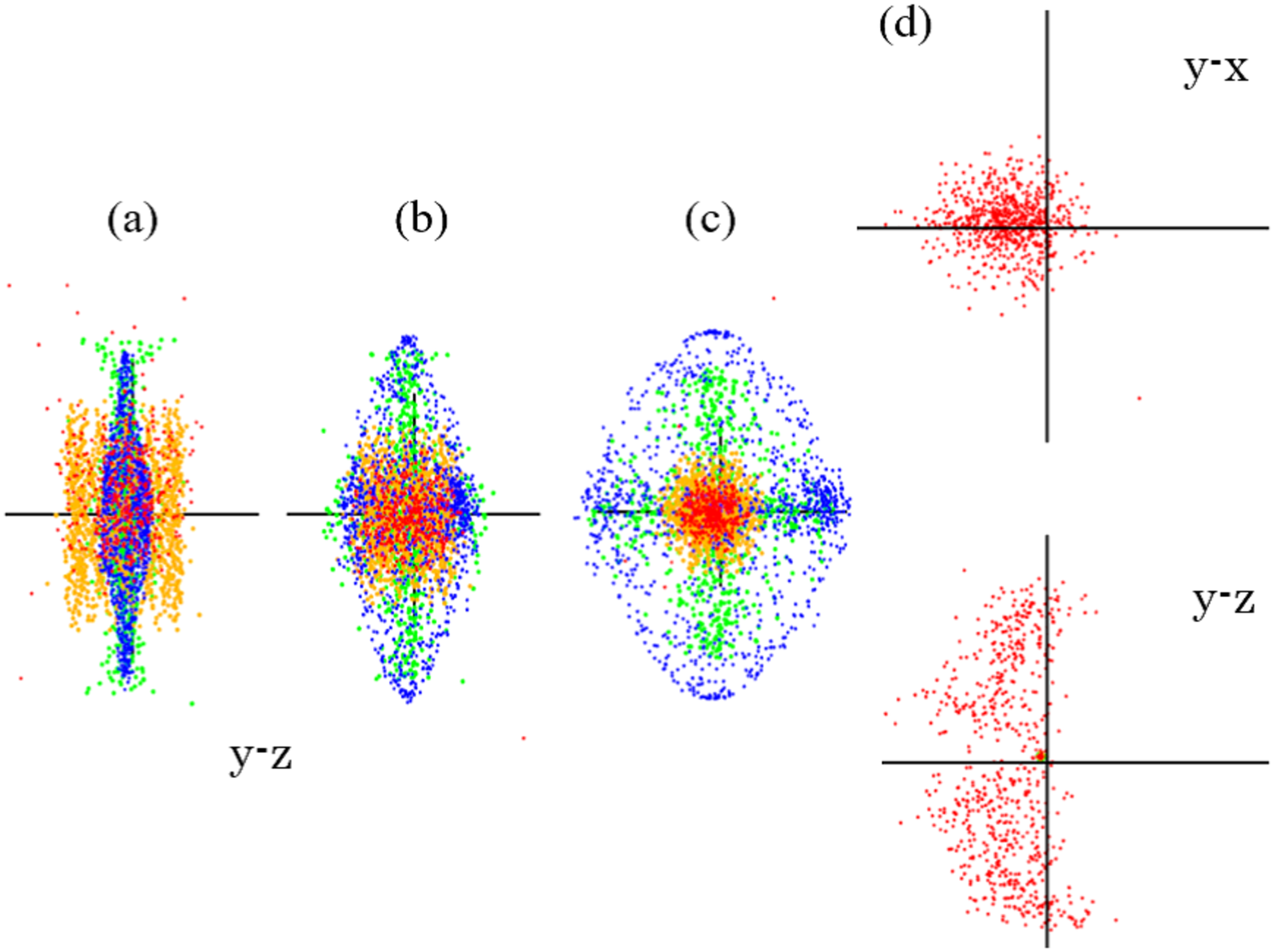} 
\caption{(Color) 
The time sequence of $5 \times 10^{17} \rm{W/cm}^2$ (Run A1) 
for enlarged $yz$ plots at (a) $13\rm{fs}$, (b) $20\rm{fs}$, (c) $45\rm{fs}$, 
and (d) for the reduced $yx$ and $yz$ plots, mostly of electrons, 
in $45\rm{fs}$. 
Protons are shown in blue, carbons by green, golds by gold,
and electrons in red.
The scales are $3.3 \times 10^{-6}\rm{cm}$, $5.5 \times 10^{-6}\rm{cm}$, 
$1.6 \times 10^{-5}\rm{cm}$ in (a) to (c), respectively, and
$9.4 \times 10^{-4}\rm{cm}$ in two vertical plots (d). }
\label{RunA1}
\end{figure}

\begin{figure} 
\centering
\includegraphics[width=8.0cm,clip]{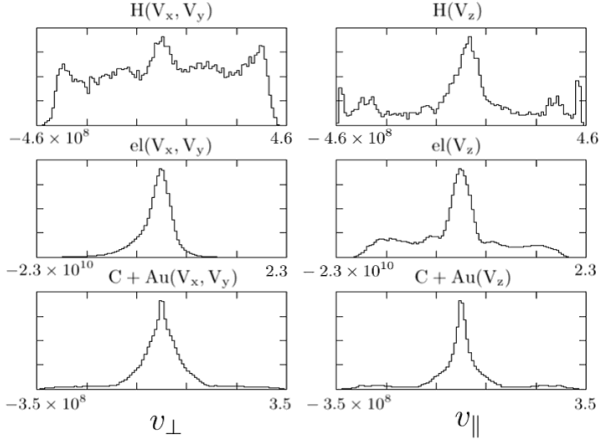} 
\caption{
The velocity distribution functions of protons, electrons, and carbon 
plus gold ions for Run A1 (from the top to bottom panels), respectively.
A time is 40 fs for the perpendicular direction $(V_{x}, V_{y})$ (left),
and the parallel direction $V_{z}$ (right), with c the speed of
light in $\rm{cm/s}$.
Scales are $4.6 \times 10^{8}$, $2.3 \times 10^{10}$, and $3.5 \times 10^{8}\rm{cm/s}$ from protons, electrons, and carbons plus golds, respectively, and ordinates in linear scales are arbitrary. }
\label{velocity-A1}
\end{figure}

\section{\label{sec3}Carbon nanotube accelerator assisted by gold ions}
\subsection{\label{sec3.1}Time developments of ions and electrons}

Table \ref{Table-1} shows the results for Run A1 to A4, which have a laser intensity of $10^{16}$--$10^{22}\rm{W/cm}^2$. Each run is initiated with a Gaussian pulse at the leftward position $y=-3.5 \times 800$ nm. A laser shot radiates the nanotube sideways from the lateral direction. 

Figure \ref{RunA1} depicts the laser intensity of $5 \times 10^{17}\rm{W}/cm ^2$ for Run A1 with a time development of nanotube expansion of (a) 13 fs, (b) 20 fs, and (c) 45 fs for the $yz$ plots, and (d) the reduced $yx$ and $yz$ plots with the electrons at 45 fs.
Laser irradiation induces a sudden rapid expansion of all ions. The antecedent protons (blue) are accelerated in the parallel ($z$) and perpendicular directions of the nanotube cylinder, as seen in panels (a) - (c). The carbon (green) and gold ions (gold) are chasing protons like guard spears, and the acceleration of positive ions lasts for $t>$ 45 fs. 
On the other hand, as the electron mass is quite light $m_{e} \ll m_{i}$ (red), the electrons move around 100 times more for a larger space than the positive ions. 
The electrons are spread wider than the ions, but they are limited to $10 \rm{\mu m}$ and have a double-humped $yz$ profile with $y \le 0$ in the panel (d). 

Figure \ref{velocity-A1} depicts the velocity distribution functions for Run A1. There is a main central peak which extends with a tail formation for protons and electrons at $5 \times 10^{17}\rm{W/cm}^2$. The protons have positive and negative tails in $V_{\parallel}$ and $V_{\perp}$ directions. These tails are somewhat higher in the $V_{\perp}$ direction. Carbon and gold ions have only thin tails, whereas electrons have thick tails in the both $V_{\parallel}$ directions.

\begin{figure} 
\centering
\includegraphics[width=9.0cm,clip]{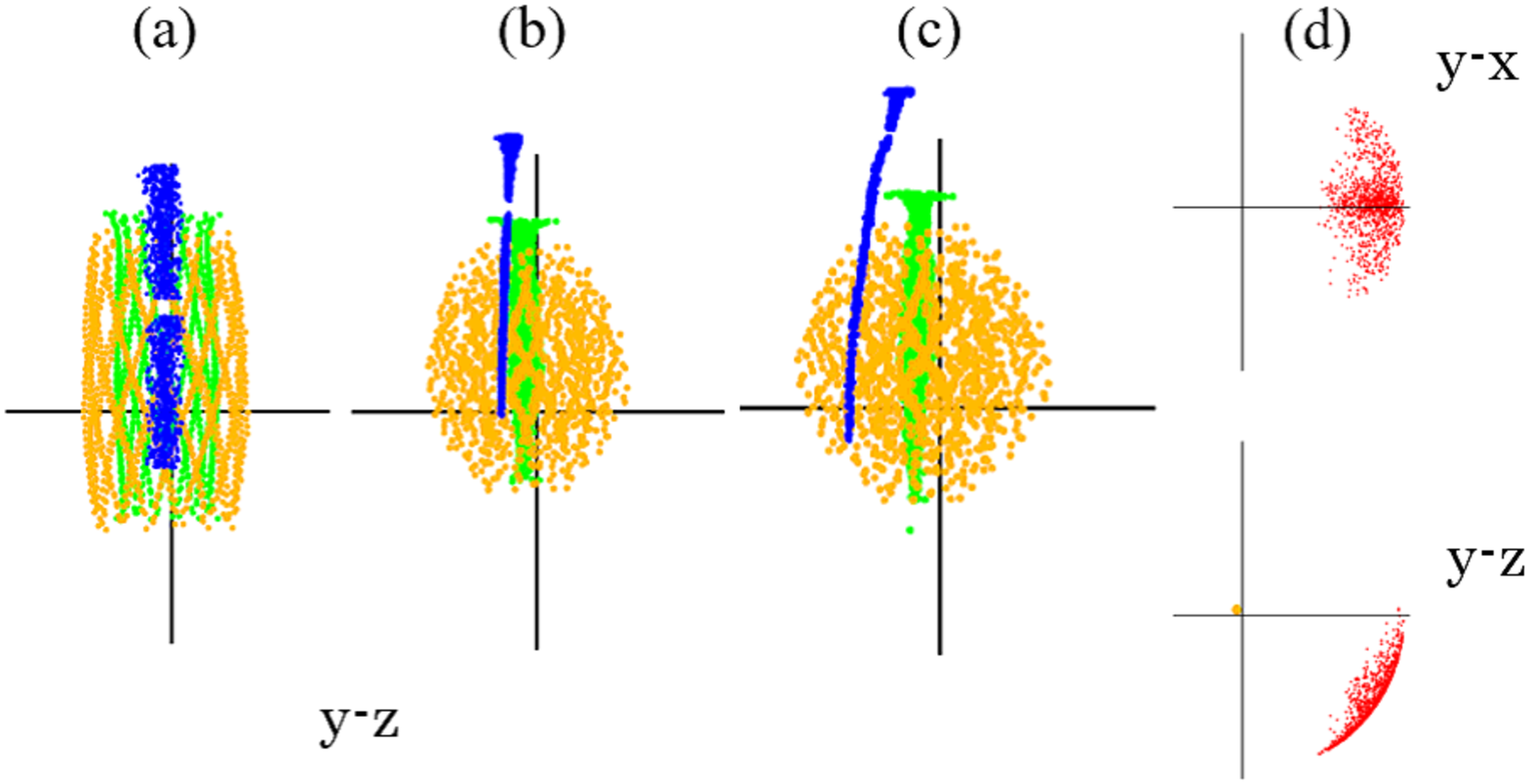}  
\caption{(Color) 
The time sequence of $1 \times 10^{22} \rm{W/cm}^2$ (Run A4) 
for enlarged $yz$ plots at (a) $3\rm{fs}$, (b) $10\rm{fs}$, (c) $21\rm{fs}$, 
and (d) for reduced $yx$ and $yz$ plots, mainly of electrons, at
$21\rm{fs}$.
Protons are shown in blue, carbons by green, golds by gold,
and electrons in red.
The scales are $3.5 \times 10^{-6}\rm{cm}$, $1.2 \times 10^{-5}\rm{cm}$ and
$2.6 \times 10^{-5}\rm{cm}$ in (a) to (c), respectively, and 
$6.2 \times 10^{-4}\rm{cm}$ in two vertical plots (d).}
\label{RunA4}
\end{figure}

\begin{figure} 
\centering
\includegraphics[width=8.0cm,clip]{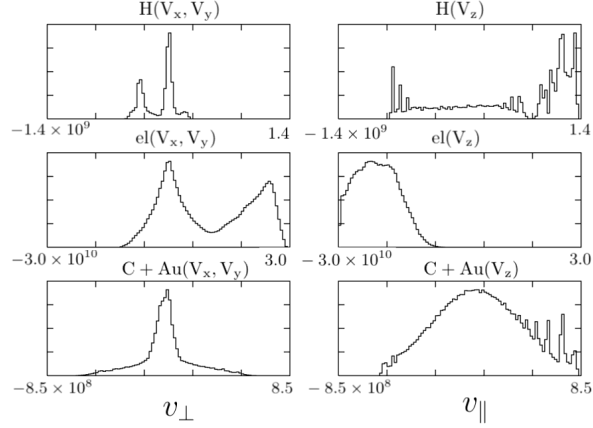} 
\caption{
The velocity distribution functions of protons, electrons, and carbon 
plus gold ions for Run A4 (from the top to bottom panels), respectively.
A time is 21 fs for the perpendicular direction $(V_{x}, V_{y})$ (left),
and the parallel direction $V_{z}$ (right), with c the speed of
light in $\rm{cm/s}$.
Scales are $1.4 \times 10^{9}$, $3.0 \times 10^{10}$, and $8.5 \times 10^{8}\rm{cm/s}$ from protons, electrons, and carbons plus golds, respectively, and ordinates in linear scales are arbitrary.}
\label{velocity-A4}
\end{figure}

\begin{figure} 
\centering
\includegraphics[width=11.0cm,clip]{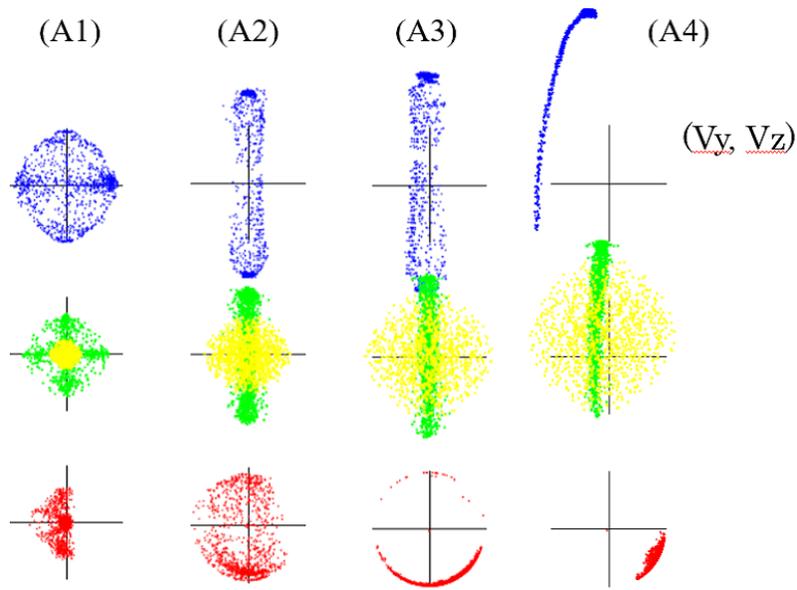} 
\caption{(Color) 
The 2D velocity plots of (i) protons, (ii) carbons and golds, and (iii) electrons, 
from the top to bottom panels, respectively.
The times for Run A1 to A4 are, (a) $45 \rm{fs}$, (b) $35 \rm{fs}$, 
(c) $35 \rm{fs}$, and (d) $21 \rm{fs}$.
The plot scales are, 100 times in the top row (for protons, blue), 
50 times in the middle row (for carbon, green, and gold, gold), and 
1.5 times for the bottom row (for electrons, red). }
\label{2D-velocity}
\end{figure}

The electron has a rest energy of 0.51 MeV~\cite{fermi}, whose kinetic energy of the same amount may be declared in the entry of the relativistic regime. 
It is $1 \times 10^{19}\rm{W/cm}^{2}$ for the carbon and gold accelerator, listed in Table \ref{Table-1}. 
At $1.7 \times 10^{19}\rm{W/cm}^2$ (Run A2), the order of energy is the first gold ions which follow carbons and electrons in the geometry space. The electrons develop a spiral shape outside the positive ions, and occupy both positive and negative $z$ coordinates. 
It will be verified by distributions of the ions to the parallel direction in Fig.\ref{2D-velocity}.

At $6 \times 10^{20}\rm{W/cm}^2$ (Run A3), a large spiral, which resembles a semi-arc shape, occupies the negative $z$ direction. The gold ion peak increases to 12 MeV, 
and the electrons have a second energy position at 6 MeV; 
carbon ions fall down to the third energy position 0.68 MeV. Regardless of the order, the carbon ions are the most important component because the six-membered cage of the nanotube must be supported initially.

At $1\times 10^{22}\rm{W/cm}^2$, all the ions and electrons flare up in massive electromagnetic radiation in Fig.\ref{RunA4} (Run A4). 
A simulation with meshes of $|x| \le 500$ \ \AA \ and $|z| \le 750$\ \AA \ ($2.5 \AA$ intervals) covers the rectangular meshes of the electromagnetic interactions. 
Fast pulsation oscillations occur for the gold and carbon ions in the phase $t \cong 6-10$ fs. The maximum energy is 42 MeV for golds and 4.5 MeV for carbons.
A deceleration of positive golds and carbons is then followed in a later stage, and the gold ions converge to 22 MeV at $t=$ 20 fs.  
The proton energy is fairly small because its mass is about $200^{-1}$ that of the gold ions.
The electrons are completely separated from all the ions, and proceed in the $E \times B$ drift to the $y>0$ direction. 
The energy of the electrons is 99\% of the total energy as early as 4 fs, and its energy becomes 37 MeV for the entire time.

The velocity distribution functions shows again the pulsation oscillations for the gold ions in Fig.\ref{velocity-A4}. 
The proton distribution function of $V_{\parallel}$ is widely distributed from the maximum positive to the negative parts. 
Protons in the parallel direction have a velocity at $1.4 \times 10^{9}\rm{cm/s}$ and are broadly distributed from the top and bottom portions, $(-1 \sim 1.4) \times 10^{9}\rm{cm/s}$. 
The carbon and gold ions in $V_{\parallel}$ have half the velocity of the protons. The population centered at $8\times 10^{8}  \rm{cm/s}$ in the parallel direction is due to the carbon ions, while that centered at $3 \times 10^{8}\rm{cm/s}$ is due to the gold ions. 
The electrons are relativistic where the $V_{y}$ decentered plot shows $(-2 \sim -3) \times 10^{10}  \rm{cm/s}$ in the middle of Fig.\ref{velocity-A4} (superpositions of the central distribution in $V_{x}$ and the decentered one in $V_{y}$ are shown). 
The decentered electron distribution was also presented in the bottom $yz$ plot of Fig.\ref{RunA4}(d).

\begin{figure} 
\centering
\includegraphics[width=10.5cm,clip]{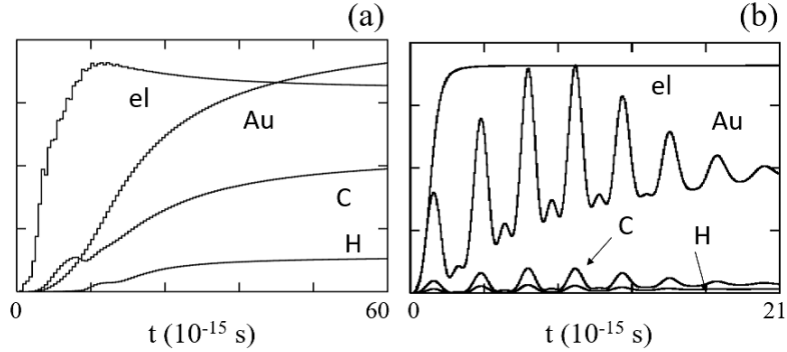}  
\caption{
The time changes of kinetic energy of golds, carbons, protons, and 
electrons for (a) $5 \times 10^{17} \rm{W/cm}^2$, and 
(b) $1 \times 10^{22} \rm{W/cm}^2$.
The maximum values are, 
(a) 0.59 MeV for ions (C,Au,H), and 0.33 MeV for electrons,
(b) 42 MeV for golds, and 37 MeV for electrons, respectively.
The final value of gold ions is 22 MeV for (b).}
\label{kinetic-en4}
\end{figure}

\subsection{\label{sec3.2}Behavior of energy and efficiency}

The momentum and energy of each particle reveal the acceleration behavior of plasmas. The momentum of the species with $N_s$ particles is defined by $W_{kin,s}= \sum_{i=1}^{N_s} (\vec{p}_{i}^2/2m_{i}) $. The total momentum determines the course of the dynamical system. 
The separate behavior of the ions or electrons is highlighted, which is divided by $N_{s}$ as $w_{kin,s}= \sum_{i=1}^{N_s} (\vec{p}_{i}^2/2m_{i})/N_{s}$. 

The energies of the golds, carbons, and protons increase monotonically in Run A1 of Fig.\ref{kinetic-en4}(a). The gold ions grow twice as fast as the carbon ions at 45 fs, and the electric interactions with these heavy ions are important. However, the growth of the protons is one order of magnitude smaller than that of heavy ions. 
The electrons become saturated around $t=$ 12 fs, but their growth is still half of that for the gold ions for Run A1. The energies on the average are 0.55 MeV for golds, 0.30 MeV for carbons, 0.082 MeV for protons, and 0.33 MeV for electrons. 
%
Figure \ref{kinetic-en4}(b) shows the time change of the energy for Run A4 at $\Delta z=2.5 \ \AA$. 
The maximum energy is 42 MeV for golds as in Fig.\ref{RunA4}.
On the other hand, the $\Delta z=2.5 \ \AA$ spacing for Run A3 is almost the same consequences in the entire regime with the $\Delta z=5 \ \AA$ run.
%

Figure \ref{2D-velocity} compares the velocity distribution functions in the $(V_{y},V_{z})$ space for Runs A1 - A4. 
For Run A1, the protons, carbons, and golds are symmetric in $(V_{y},V_{z})$ of the first two rows of panel (a). The electrons in the third row occupy mainly the leftward hemisphere $V_{y}$, and both the top and bottom parts resemble a butterfly pattern in the $V_{z}$ direction. 
Run A2 in panel (b) is symmetric for the positive ions and the distributions are straight up toward the parallel direction. The electrons form a large arc in the entire $V_{y}-V_{z}$ space. 
The energy of $10^{19}\rm{W/cm}^{2}$ is the beginning of the relativistic regime, as panel (b) of Fig.\ref{2D-velocity}.
For Run A3 in (c), the electrons form an antisymmetric arc, which appears quite large mostly in the negative regime, $V_{z} <0$. 
Finally, for Run A4 in (d), the pipe shape for the protons is totally antisymmetric, and the electrons exhibit a very large island in the fourth quadrant with $V_{y}>0$ and $V_{z}<0$.

The momentum elucidates the energy efficiency. 
The momenta of Run A1 are 17\%, 32\%, 0.86\%, and 50\% for carbons, golds, protons,  
and electrons, respectively. All ions and electrons are almost the same amount divided, and the ratio of golds to electrons is $\rm{(Au/electron)}=$ 64\%. 
As the laser intensity increases from Run A2 to Run A3, the energy efficiency of the ions drops exponentially. The energy efficiency of the golds to electrons for Run A3 becomes 3.1\%. 
At the large intensity Run A4, the ratios are $ 6 \times 10^{-2}$\%, 0.82\%, $5 \times 10^{-3}$\% and 99\% for carbons, golds, protons, and electrons, respectively. The ratio of golds to electrons is $\rm{(Au/electron)}=$ 0.65\%.

The trend for the kinetic energy over the intensity $>10^{20}\rm{W/cm}^2$  shows that the gold ions are approaching parabolic sub-relativistic in Table \ref{Table-1}. It is compared to a linear relativistic regime for the electrons.

\begin{table} 
\caption{The carbon, gold and electron energies (MeV, average), and energy efficiency 
of golds to electrons (\%) for the number of gold layers within a nanotube. 
Quotation to Run A4 for the single cage (with protons) is shown as the first line. }
\label{Table-2}
%
  \centering 
  \vspace{0.1cm}
  \begin{tabular}{cccccc} \toprule 
    Series & layers & C & Au & electron & Efficiency \hspace{0.3cm} \\ \hline 
    A4      & 1	 & 1.9 & 21.9 & 37.3 & 0.78\% \hspace{0.3cm} \\ 
    B1      & 2   & 2.0 & 23.5 & 29.6 & 0.46\% \hspace{0.3cm} \\ 
    B2      & 3   & 2.3 & 29.1 & 50.8 & 0.98\% \hspace{0.3cm} \\
  \toprule
  \end{tabular}
\end{table}

\begin{table}
\caption{Three aligned nanotube cages in the $y$ and $x$ direction for Run C1 and Run C2, respectively, where two layers are fabricated inside of the nanotube.}
\label{Table-3}

\vspace{0.2cm}
\centering
\begin{tabular}{cccccc} \toprule 
    Series & layers & C & Au & electron & efficiency \hspace{0.3cm} \\ \midrule
    C1 & y1,y2,y3 & 3.3	& 34.3	& 46.2 & 0.98\% \hspace{0.3cm} \\ \midrule
    C2	& x1,x2,x3 & 3.2 & 34.1 & 44.6 & 1.00\% \hspace{0.3cm} \\ \bottomrule
\end{tabular}
\end{table}

\begin{figure} 
\centering
\includegraphics[width=7.50cm,clip]{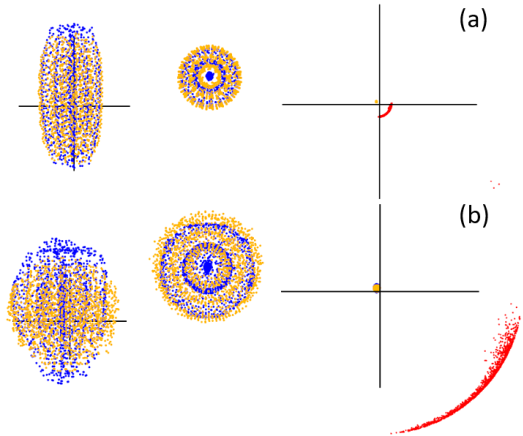} 
\caption{(Color) The cylindrical nanotube for a triple cages at $1 \times 10^{22}\rm{W/cm}^2$ in (a) 3 fs and (b) 25 fs, from the top and bottom rows, respectively.
The enlarged plots are the $yz$ and $yx$ directions for positive ions, respectively, and the $yz$ direction for electrons, with carbons (blue), golds (gold), and electrons (red). 
The final value for the triple-stacked cylindrical cage is 29.1 MeV for gold ions and 50.8 MeV for electrons.
The scales for positive ions are, $3.1 \times 10^{-6} \rm{cm}$ and $2.3 \times10^{-5} \rm{cm}$ for (a) and (b), respectively, and $7 \times 10^{-4} \rm{cm}$ for electrons.}
\label{merging}
\end{figure}

\begin{figure}
\centering
\includegraphics[width=9.00cm,clip]{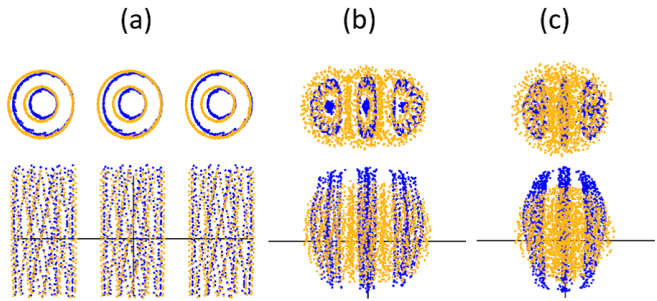}
\caption{(Color) 
Merging of a three-array nanotube with double layers inside for (a) 1 fs, (b) 7 fs, and (c) 24 fs, and in carbon (blue) and gold (gold) ions. The scales in the $yx$(top) and $yz$(bottom) directions are, $1.3 \times 10^{-6} \rm{cm}$, $6.3 \times 10^{-6} \rm{cm}$, $2.5 \times 10^{-5} \rm{cm}$ for (a) to (c), respectively.}
\label{merging2}
\end{figure}

While in the cylindrical nanotube of the single cage in Run A, it is built up for a double or triple cages for $1 \times 10^{22}\rm{W/cm}^2$ in Tables \ref{Table-2} of Run B. The $E \times B$ drift speed remains the same as before. 
Rapid fast oscillations of gold ions for a triple-stacked cage are developed in Run B2 of Fig.\ref{merging}. The energy at the final time at $t=$ 30 fs  is about 80\% that of the maximum value at $t \cong$ 12 fs.
The energy of the double-stacked cylindrical cage at the final time is 23.5 MeV for gold ions in Run B1 (on the average), and that of the triple-stacked cylindrical cage is 29.1 MeV for Run B2. The energy efficiency for the double and triple-stacked cages is $\rm{(Au/electron)}=$ 0.46\% and 0.98\%, respectively.

A multi-nanotube consisting of three straight arrays with double layers inside the tube is shown for Run C1 and Run C2, respectively, in $y$ and $x$ direction in Table \ref{Table-3}. 
The cages are compressed with each other and become fat as a time develops in Fig.\ref{merging2}. The positive ions are strongly nested with other nanotubes at $t=$ 24 fs.  
Electrons in the distribution function make a semi-arc shape in the fourth quadrant, $V_{y}<0, V_{z}>0$.  
The energy is improved for the gold ions to be 34.1 MeV and 34.3 MeV for (a) and (b) of Table \ref{Table-3}, respectively. The maximum efficiency is $\rm{(Au/electron)}=$ 1.0 \% for Run C2.

\section{\label{sec4}Summary}
 
Relativistic and electromagnetic molecular dynamics was described for ultra-high temperature and a few tens of MeV behaviors.  Highly accurate interactions were executed to real charge and mass of particles. The Coulomb potential field was collected in the infinite space, and the electromagnetic fields were calculated in the coordinate space.
The transverse electric field and the Coulomb forces were used in the equation of motion, which was a good approximation  because the physical process was contained in 100 nm of the accelerated nanotube. The light travels over one-tenth of the period of an 800-nm (2.7fs per period) sapphire laser.

As the relativistic molecular dynamics simulation, the nanotube accelerator was performed using the Coulomb potential and electromagnetic fields. 
With the $E \times B$ pulse of 800 nm, the positive ions were accelerated in homogeneous directions in the $5 \times 10^{17}\rm{W/cm}^{2}$ range. 
Exceeding the $10^{18}\rm{W/cm}^{2}$ range, the relativistic regime appeared clearly and the ions were straight in the parallel direction, whereas electrons were asymmetric and proceeded in the oblique direction.
At the large intensity $10^{22}\rm{W/cm}^{2}$, the massive electromagnetic radiation for short ranges and the rapid expansion to the infinite space were present in the simulation. The gold and carbon ions were flared up at pulsation oscillations, and electrons acquired 99\% of the speed of light. 
Simulations of multi-nanotubes were executed at $1 \times 10^{22}\rm{W/cm}^2$, and the triple-stacked cylindrical cage was 29 MeV.
Three arrays of a multi-nanotube were shown to be 34 MeV for gold ions, and the energy efficiency was 1\%. 

The summary is that the numerical simulation method was presented for ultra-high temperature and a few tens of MeV behaviors. The simulation code was utilized in Coulomb potential and electromagnetic fields, which was examined for resolution accuracy.
It was the cooperation of Coulomb potential and electromagnetic radiation of plasmas by the relativistic molecular dynamics simulation.

\newpage

\noindent {\bf 
Acknowledgments}\\

One of the authors (M.T.) would like to thank Professor A. Iiyoshi, Professor M. Sato and Professor Y. Zempo for kindness, and physics and computational discussions.
The Principal Editor Professor D. W. Walker and the Specialist Editor Professor K. Germaschewski are deeply acknowledged.
The present simulations were performed using Fujitsu FX100 Supercomputer at the National Institute for Fusion Science, Japan.


    \renewcommand{\theequation}{A.\arabic{equation}}
    \setcounter{equation}{0}

\vspace{1.0cm}
\noindent
{\bf Appendix A: Motion and Maxwell Field Equations by the International System of Units}
\label{append-A}

\vspace{0.5cm}
The equation of motion for plasmas is written in the international system of units as,  
\begin{eqnarray}
d \vec{p}_{i}/dt &=& - \nabla \sum_{j=1}^{N} [q_{i}q_{j} /4 \pi \epsilon_{0} r_{ij} + \Phi (r_{i},r_{j})] 
 \nonumber \\[-0.2cm]
 && \hspace{1.3cm} 
 + q_{i} [\vec{E}_{T}(\vec{r}_{i},t) +\vec{v}_{i} \times \vec{B}(\vec{r}_{i},t)],
\label{eq_motionM} \\
d \vec{r}_{i}/dt &=& \vec{v}_{i}, \ \ 
\vec{p}_{i}= m_{i}\vec{v}_{i}/\sqrt{1 - (\vec{v}_{i}/c)^2}.
\end{eqnarray}
The Coulomb potential and transverse electric fields are utilized in this code.
The $\Phi(r_{i},r_{j})$ is the Lennard-Jones potential, the $\vec{E}_{T}$ and $\vec{B}$ are the transverse electric and magnetic fields, respectively, and $\mu_{0}= 4 \pi \times 10^{-7} \rm{W/Am}$ and $\epsilon_{0}= 1/\mu_{0}c^2$. 
%
The Maxwell field equations are written as,
\begin{eqnarray}
\nabla \cdot \vec{E}(\vec{r}, t) &=& (1/\epsilon_{0}) \sum_{i=1}^{N} q_{i}S(\vec{r} -\vec{r_{i}}),\\
\partial \vec{B}/\partial t &=& - \nabla \times \vec{E}
\label{ddd.eqM}, \\
\mu_{0} \epsilon_{0} \ \partial \vec{E}/\partial t &=& \nabla \times \vec{B}  - \mu_{0} \sum_{i=1}^{N} q_{i} \vec{v}_{i} S(\vec{r}-\vec{r_{i}}),  
\label{Ampere.eq} \\
\nabla \cdot \vec{B}= 0. 
\end{eqnarray}
The magnetic field in plasmas is $\vec{B}= \mu_{0}\vec{H}$. The symbol $L$ is the longitudinal electric field, and the shape function $S(\vec{r})$ is prorated to nearest grids.
The values are, $c= 2.998 \times 10^{8} \rm{m/s}$, $e= 1.602 \times 10^{-19}\rm{C}$, 
$m_{e}= 9.111 \times 10^{-31}\rm{kg}$, and $1\rm{eV}= 1.602 \times 10^{-19}\rm{J}$.

\newpage

\noindent
{\bf Appendix B: Parallelization in Coulomb Forces and Electromagnetic Fields}
\label{append-B}

\vspace{0.5cm}
The molecular dynamics simulation runs for the Coulomb forces and electromagnetic field calculations. 
The Coulomb field of $N \times N$ particle summation is parallelized in the utilization ratio 94 \% at 100 ranks, where the ranks are twice the total processors.
On average, it takes 1.20 s per time step for the Coulomb forces (Run A4) of 52 ranks by 16 processors using the OpenMP technique by Fujitsu FX100 Supercomputer. 
The Maxwell equations are solved in Eqs.(\ref{ddd.eq})-(\ref{ddd.divB}), the finite grid correction Eqs.(\ref{ETcorre1}),(\ref{ETcorre2}), and the field splitting Eqs.(\ref{EL1}),(\ref{EL2}), as written in Sec.\ref{sec2.1}.
The computation of the Coulomb forces and redundant electromagnetic field calculations is 5.12 s per time step with the memory of 13 GB per processor ($200^{2} \times 600$ meshes, $\Delta x= 2.5 \ \AA$).

The electromagnetic fields in the coordinate space is parallelized by domain decomposition if a large number of processors are executed. Different fields are allocated in a minimal overlap of adjacent processors.
One-dimensional ($z$) field arrays are handled by $mpi{\_}isend$ and $mpi{\_}irecv$ subroutines where only the top (or bottom) processor is connected to the inside region due to non-periodic boundary conditions. 
The execution of the parallel code takes 3.97 s per time step for the 52 ranks, with the Coulomb forces of 1.40 s per time step included, with the memory of 11 GB per processor (each of $200^{2} \times 12$ meshes with $\Delta x= 2.5 \ \AA$). 
%

\newpage

\end{document}